\documentclass[10pt,letterpaper]{article}
\usepackage[top=0.85in,left=2in,footskip=0.75in,marginparwidth=2in]{geometry}

\pdfoutput=1

\usepackage[utf8]{inputenc}
\DeclareUnicodeCharacter{2010}{-}

\usepackage{cite}

\usepackage{nameref}
\usepackage{hyperref}

\usepackage[right]{lineno}

\usepackage{microtype}
\DisableLigatures[f]{encoding = *, family = * }

\raggedright
\usepackage{changepage}

\usepackage[aboveskip=1pt,labelfont=bf,labelsep=period,singlelinecheck=off]{caption}

\makeatletter
\renewcommand{\@biblabel}[1]{\quad#1.}
\makeatother

\usepackage{lastpage,fancyhdr,graphicx}
\usepackage{epstopdf}
\fancyheadoffset[L]{2.25in}
\fancyfootoffset[L]{2.25in}

\usepackage{color}

\definecolor{Gray}{gray}{.25}

\usepackage{graphicx}

\usepackage{sidecap}

\usepackage{wrapfig}
\usepackage[pscoord]{eso-pic}
\usepackage[fulladjust]{marginnote}
\reversemarginpar
\usepackage{nameref}
\usepackage{color,soul}

\makeatletter
\AtBeginDocument{\let\hl\@firstofone}
\makeatother

\begin{document}

\vspace*{0.35in}

\begin{flushleft}
{\Large
\textbf{Centralized and distributed cognitive task processing in the human connectome}
}
\newline
\\
Enrico Amico \textsuperscript{1}\textsuperscript{2},
Alex Arenas \textsuperscript{3} and
Joaqu\'{i}n Go\~{n}i \textsuperscript{1}\textsuperscript{2}\textsuperscript{4}*
\\
\bigskip
\small
\ 1 School of Industrial Engineering, Purdue University, West-Lafayette, IN, USA
\\
\ 2 Purdue Institute for Integrative Neuroscience, Purdue University, West-Lafayette, IN, USA
\\
\ 3 Departament d'Enginyeria Inform\`{a}tica i Matem\`{a}tiques, Universitat Rovira i Virgili, Tarragona, Spain
\\
\ 4 Weldon School of Biomedical Engineering, Purdue University, West-Lafayette, IN, USA
\\
\bigskip
* jgonicor@purdue.edu

\end{flushleft}
\justifying

\begin{abstract}
A key question in modern neuroscience is how cognitive changes in a human brain can be quantified and captured by functional connectomes (FC) . A systematic approach to measure pairwise functional distance at different brain states is lacking. This would provide a straight-forward way to quantify differences in cognitive processing across tasks; also, it would help in relating these differences in task-based FCs to the underlying structural network. Here we propose a framework, based on the concept of Jensen-Shannon divergence, to map the task-rest connectivity distance between tasks and resting-state FC. We show how this information theoretical measure allows for quantifying connectivity changes in distributed and centralized processing in functional networks. We study resting-state and seven tasks from the Human Connectome Project dataset to obtain the most distant links across tasks. We investigate how these changes are associated to different functional brain networks, and use the proposed measure to infer changes in the information processing regimes. Furthermore, we show how the FC distance from resting state is shaped by structural connectivity, and to what extent this relationship depends on the task. This framework provides a well grounded mathematical quantification of connectivity changes associated to cognitive processing in large-scale brain networks.
\end{abstract}

\section*{Introduction}

The progress in neuroimaging methodologies in recent years, together with the rise of publicly available datasets \cite{van_essen_human_2012,van_essen_wu-minn_2013} has boosted research on quantitative analysis of brain connectivity patterns based on network science. The intuition of modeling the brain as a network \cite{fornito_fundamentals_2016,sporns_human_2011,bullmore_complex_2009,bassett2006small} has rapidly expanded into the scientific area denominated Brain Connectomics \cite{fornito_fundamentals_2016,fornito_connectomics_2015}. In brain network models, nodes correspond to grey-matter regions (based on brain atlases or parcellations) while links or edges correspond to structural or functional connections. Structural connections are estimated from diffusion weighted imaging \cite{bihan_looking_2003,tournier_diffusion_2011} data by modeling white matter pathways through tractography algorithms \cite{tournier_mrtrix:_2012,smith_anatomically-constrained_2012,smith_sift2:_2015}. Functional connections represent statistical dependencies between brain regions time series while subjects are either at rest or performing a task during functional MRI (fMRI) sessions \cite{van_den_heuvel_exploring_2010}. These functional associations are usually measured via correlations among fMRI time series to study functional connectivity (FC) in the human brain \cite{fox_spontaneous_2007,van_den_heuvel_exploring_2010}. 

These recent advances have led the brain connectivity community to start exploring and quantify differences between resting-state FCs and task-based FCs \cite{cole_intrinsic_2014,cole_cognitive_2007,cole_multi-task_2013,gonzalez2012whole,krienen2014reconfigurable,gratton2016evidence}. \hl{Some of} the main lines of research in this direction involved: whole-brain network similarity analyses on the intrinsic and task-evoked network architecture of human connectome \cite{cole_intrinsic_2014}; the mapping of cortical hubs and brain region for adaptive task control (so called ``cognitive control network'' \cite{cole_cognitive_2007,cole_multi-task_2013}; the investigation of activity flow from resting-state FCs to infer brain regions that carry diverse cognitive task information \cite{cole_multi-task_2013,ito_cognitive_2017}. Despite all these efforts in trying to characterize connectivity differences between resting-state and task activity in brain networks, a systematic analysis on how to measure pairwise (i.e. at the level of FC links) ``cognitive distance'' between these different functional states is still lacking. Such a methodology would provide a straight-forward way to quantify differences in cognitive processing across tasks; also, it would help in relating these local differences in task-based FCs to the underlying structural network architecture, another exciting avenue for the brain connectomics community \hl{(\mbox{\cite{fukushima2018structure,hermundstad2013structural,hermundstad2014structurally,mivsic2016network,medaglia2018functional} }}, see also our recent work \cite{amico_mapping_2017-1}).   

Here we propose a framework, based on Jensen-Shannon (JS) divergence \cite{briet_properties_2009,cover_elements_2012}, to map the ``cognitive distance'' between task and resting-state functional connections. We show how this simple measure allows for quantifying the amount of changes in distributed and centralized processing in human functional networks. 

We use resting-state and seven different task sessions from the Human Connectome Project (HCP) database to obtain the most JS-distant edges across tasks. We study how these changes across tasks are associated to different functional brain networks, and use the proposed measure to infer modifications in the information processing regimes of these networks. Furthermore, we show how cognitive distance is shaped by the brain structural architecture and the level of nestedness of axonal pathways, and \hl{to what extent this relationship depends on the task}. We conclude by discussing the new insights offered by this approach, as well as potential applications and future directions.  
\section*{Methods} \label{methods}
\subsubsection*{Dataset}
The fMRI dataset used in this work is from the publicly available Human Connectome Project (HCP, \url{http://www.humanconnectome.org/}), Release Q3. Per HCP protocol, all subjects gave written informed consent to the Human Connectome Project consortium. Below is the full description of the acquisition protocol and processing steps. 
\subsubsection*{HCP: functional data}
\hl{We assessed the 100 unrelated subjects (54 females, 46 males, mean age = 29.1 $\pm$ 3.7) as provided at the HCP 900 subjects data release \mbox{\cite{van_essen_human_2012,van_essen_wu-minn_2013}}. This subset of subjects provided by HCP ensures that they are not family relatives. This criterion was crucial to exclude the need of family-structure co-variables in our analyses as well as possible identifiability confounds.}
The fMRI resting-state runs were acquired in separate sessions on two different days (HCP filenames: rfMRI{\_}REST1 and rfMRI{\_}REST2), with two different acquisitions (left to right or LR and right to left or RL) per day \cite{glasser_minimal_2013,smith_resting-state_2013}. The seven fMRI tasks were: gambling (tfMRI{\_}GAMBLING), relational (tfMRI{\_}RELATIONAL), social (tfMRI{\_}SOCIAL), working memory (tfMRI{\_}WM), motor (tfMRI{\_}MOTOR), language (tfMRI{\_}LANGUAGE, including both a story-listening and arithmetic task) and emotion (tfMRI{\_}EMOTION). The working memory, gambling and motor task were acquired on the first day, and the other tasks were acquired on the second day \cite{barch_function_2013,van_essen_wu-minn_2013}. The HCP scanning protocol was approved by the local Institutional Review Board at Washington University in St. Louis. For all sessions, data from both the left-right (LR) and right-left (RL) phase-encoding runs were averaged to calculate connectivity matrices. This operation was done for all 7 fMRI tasks. Full details on the HCP dataset have been published previously \cite{glasser_minimal_2013,smith_resting-state_2013,van_essen_human_2012}. 

\subsubsection*{HCP: structural data} 

We used DWI runs from the same $100$ unrelated subjects of the HCP 900 subjects data release \cite{van_essen_human_2012,van_essen_wu-minn_2013}. The diffusion acquisition protocol is covered in detail elsewhere \cite{glasser_minimal_2013,sotiropoulos_advances_2013,ugurbil_pushing_2013}. Below we mention the main characteristics. Very high-resolution acquisitions ($1.25$ mm isotropic) were obtained by using a Stejskal–Tanner (monopolar) \cite{stejskal_spin_1965} diffusion-encoding scheme. Sampling in q-space was performed by including $3$ shells at $b=1000$, $2000$ and $3000$ s/mm2. For each shell corresponding to $90$ diffusion gradient directions and 5 $b=0$ acquired twice were obtained, with the phase encoding direction reversed for each pair (i.e. LR and RL pairs). Directions were optimized within and across shells (i.e. staggered) to maximize angular coverage using the approach of \cite{caruyer_optimal_2011}(http://www-sop.inria.fr/members/Emmanuel.Caruyer/q-space-sampling.php), and form a total of $270$ non-collinear directions for each PE direction. Correction for EPI and eddy current-induced distortions in the diffusion data was based on manipulation of the acquisitions so that a given distortion manifests itself differently in different images \cite{andersson_how_2003}. To ensure better correspondence between the phase-encoding reversed pairs, the whole set of diffusion-weighted (DW) volumes is acquired in six separate series. These series were grouped into three pairs, and within each pair the two series contained the same DW directions but with reversed phase-encoding (i.e. a series of $M_i$ DW volumes with RL phase-encoding is followed by a series of $M_i$ volumes with LR phase-encoding, where $i = [1,2,3]$).

\subsubsection*{Brain atlas}
\label{brain atlas}
We employed a cortical parcellation into $360$ brain regions as recently proposed by Glasser et al. \cite{glasser_multi-modal_2016}. For completeness, $14$ sub-cortical regions were added, as provided by the HCP release (filename “Atlas{\_}ROI2.nii.gz”). To do so, this file was converted from NIFTI to CIFTI format by using the HCP workbench software \cite{glasser_minimal_2013,marcus_informatics_2011} (\url{http://www.humanconnectome.org/software/connectome-workbench.html}, command -cifti-create-label).  

\subsubsection*{HCP preprocessing: functional data} \label{functional data}

The HCP functional preprocessing pipeline \cite{glasser_minimal_2013,smith_resting-state_2013} was used for the employed dataset. This pipeline included artifact removal, motion correction and registration to standard space. Full details on the pipeline can be found in \cite{glasser_minimal_2013,smith_resting-state_2013}. The main steps were: spatial (“minimal”) pre-processing, in both volumetric and grayordinate forms (i.e., where brain locations are stored as surface vertices \cite{smith_resting-state_2013}); weak highpass temporal filtering ($>2000$s full width at half maximum) applied to both forms, achieving slow drift removal. MELODIC ICA \cite{jenkinson_fsl_2012} applied to volumetric data; artifact components identified using FIX \cite{salimi-khorshidi_automatic_2014}. Artifacts and motion-related time courses were regressed out (i.e. the $6$ rigid-body parameter time-series, their backwards-looking temporal derivatives, plus all $12$ resulting regressors squared)  of both volumetric and grayordinate data \cite{smith_resting-state_2013}. 

For the resting-state fMRI data, we also added the following steps: global gray matter signal was regressed out of the voxel time courses \cite{power_methods_2014}; a bandpass first-order Butterworth filter in forward and reverse directions [$0.001$ Hz, $0.08$ Hz] \cite{power_methods_2014} was applied (Matlab functions \textit{butter} and \textit{filtfilt}); the voxel time courses were z-scored and then averaged per brain region, excluding outlier time points outside of $3$ standard deviation from the mean, using the workbench software \ ( workbench command -cifti-parcellate ). For task fMRI data, we applied the same above mentioned steps but we opted for a more liberal bandpass filter [$0.001$ Hz, $0.25$ Hz], since it is still unclear the connection between different tasks and optimal frequency ranges \cite{cole_intrinsic_2014}.

Pearson correlation coefficients between pairs of nodal time courses were calculated (MATLAB command \textit{corr}), resulting in a symmetric connectivity matrix for each fMRI session of each subject. \hl{In this paper we will refer to this matrix as functional connectivity matrix or functional connectome (FC)}. As aforementioned, data from both the left-right (LR) and right-left (RL) phase-encoding runs were averaged to calculate individual functional connectomes in each fMRI session. Functional connectivity matrices were kept in its signed weighted form, hence neither thresholded nor binarized. Finally, the resulting individual functional connectivity matrices were ordered (rows and columns) according to $7$ functional cortical sub-networks (FNs) as proposed by Yeo and colleagues \cite{yeo_organization_2011}. \hl{To do so, for each brain region in Glasser atlas, the FN-membership (as a percentage) to each of the $7$ functional networks (sum of the membership vector being equal to $1$) was calculated. Finally, each brain region was assigned to the most highly present FN.} For completeness, an $8$th sub-network including the $14$ HCP sub-cortical regions was added  (as analogously done in recent papers \cite{amico_mapping_2017-1,amico_mapping_2017}).

\subsubsection*{HCP preprocessing: structural data}\label{structuraldata}

The HCP DWI data were processed following the MRtrix3 \cite{tournier_diffusion_2011} guidelines (\url{http://mrtrix.readthedocs.io/en/latest/tutorials/hcp_connectome.html}). In summary, we first generated a tissue-segmented image appropriate for anatomically constrained tractography (ACT \cite{smith_anatomically-constrained_2012}, MRtrix command 5ttgen); we then estimated the multi-shell multi-tissue response function (\cite{christiaens_global_2015}, MRtrix command dwi2response msmt{\_}5tt) and performed the multi-shell, multi-tissue constrained spherical deconvolution (\cite{jeurissen_multi-tissue_2014}, MRtrix dwi2fod msmt{\_}csd); afterwards, we generated the initial tractogram (MRtrix command tckgen, $10$ million streamlines, maximum tract length $= 250$, FA cutoff $= 0.06$) and applied the successor of Spherical-deconvolution Informed Filtering of Tractograms (SIFT2, \cite{smith_sift2:_2015}) methodology (MRtrix command tcksift2). Both SIFT \cite{smith_sift:_2013} and SIFT2 \cite{smith_sift2:_2015} methods provides more biologically meaningful estimates of structural connection density. SIFT2 allows for a more logically direct and computationally efficient solution to the streamlines connectivity quantification problem: by determining an appropriate cross-sectional area multiplier for each streamline rather than removing streamlines altogether, measures of fiber connectivity are obtained whilst making use of the complete streamlines reconstruction \cite{smith_sift2:_2015}. Finally, we mapped the SIFT2 outputted streamlines onto the $374$ chosen brain regions ($360$ from Glasser et al. \cite{glasser_multi-modal_2016} brain atlas plus $14$ subcortical regions, see Brain Atlas section) to produce a structural connectome (MRtrix command tck2connectome). Finally, a $\log_{10}$ transformation \cite{fornito_fundamentals_2016} was applied on the structural connectomes \hl{(SC, i.e. the anatomical networks)} to better account for differences at different magnitudes. In consequence, SC values ranged between $0$ and $5$ on this dataset.

\subsection*{Jensen-Shannon distance on functional edges}\label{jsd}
The Jensen-Shannon divergence is a method commonly used to measure dissimilarities between two probability distributions \cite{cover_elements_2012,de_domenico_structural_2015,briet_properties_2009}. In the case of two discrete probability distributions $P$ and $Q$, the Jensen-Shannon divergence ($JSD$) is defined by:
\begin{equation}
  JSD(P||Q) = \frac{1}{2}D_{KL}(P || M) + \frac{1}{2}D_{KL}(Q || M) 
\end{equation}  
where $M = \frac{1}{2} (P + Q)$ and $D_{KL}$ is the Kullback-Leibler divergence \cite{cover_elements_2012}. For two  discrete probability distributions $P$ and $Q$, it is defined by:
\begin{equation}
  D_{KL}(P||Q) = - \sum\limits_{i} P(i) \log \frac{Q(i)}{P(i)}
\end{equation}  
For the particular case of measuring the dissimilarity between two probability distributions $P$ and $Q$, the Jensen-Shannon divergence is bounded between 0 and 1, given that one uses the base 2 logarithm:
\begin{equation}
0 \leq JSD(P||Q) \leq 1 
\end{equation}

\hl{It has been shown that the square root of the Jensen-Shannon divergence is a well-defined distance metric \mbox{\cite{endres2003new,osterreicher2003new}}, often referred to as ``Jensen-Shannon distance''}:

\begin{equation}
JS_{dist}(P||Q) = \sqrt{JSD(P||Q)} 
\end{equation}

\begin{figure}[!ht] 
\centerline{\includegraphics[width=\textwidth]{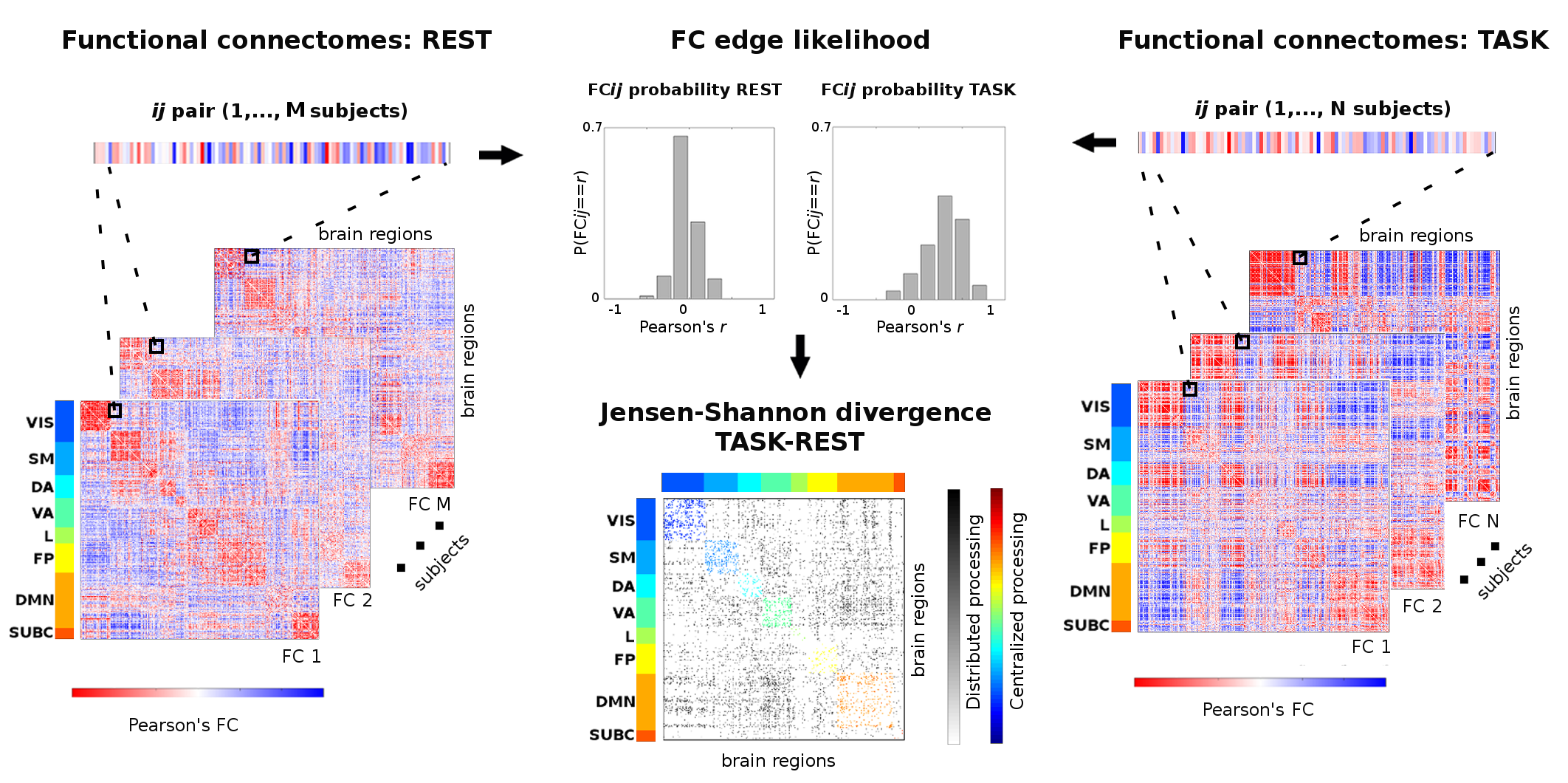}}
\caption{\textbf{Workflow scheme for task-rest connectivity distance \hl{(unpaired)}.} This scheme summarizes the procedure to measure edgewise distance from two cohorts of \hl{(M and N)} functional connectomes (FCs) at rest (left) to a task-based one (right). First, an edge ij is extracted from the set, for both the resting-state and task-based FCs; these two vectors of \hl{M and N} connectivity values are then transformed into probability distributions (center top); finally, the Jensen-Shannon distance for these two edgewise probabilities is computed (center bottom). Iterating this procedure over all possible ij pairs gives a Jensen-Shannon (JS) matrix of local distance in task FCs with respect to the REST baseline. The JS matrix is ordered by the 7 functional networks (FNs): visual (VIS), somato-motor (SM), dorsal attention (DA), ventral attention (VA), limbic (L), frontoparietal (FP), default mode network (DMN). An eighth subcortical network (SUBC) is added for completeness. Within-networks most distant edges are color-coded according FNs. Between-networks most distant edges are in gray-scale. This method allows for quantifying the changes between centralized (within-network) and distributed (between network) processing when a specific task is performed with respect to the resting-state baseline.}
\label{fig1}
\end{figure}

We used the $JS_{dist}$ to map ``connectivity distance'' between resting state and task sessions. Here we assume resting state to be the ``cognitive baseline", and we measured the $JS_{dist}$ link to link from every task FCs to resting state FCs. Below follows a detailed description of the procedure (see also scheme at Fig. \ref{fig1}). First, for every edge in a functional connectome, we extracted the corresponding individual values \hl{(out of $100$ HCP subjects, we picked 50 subjects for resting-state FCs and 50 different subjects for the task FCs)}. In this study, this resulted in having two vectors with \hl{$50$} elements, one for each resting state edge and one for each edge in the task FCs whose JS distance is to be evaluated. These vectors represent Pearson's correlation distributions of connectivity values across all subjects in the cohort. 
Secondly, we transform these two Pearson's distributions into discrete probability distributions. We sampled the $[-1,1]$ Pearson's range via uniform binning (bin width $= 0.2$), and counted the likelihood of occurrence of the connectivity values in each bin. 
Finally, the $JS_{dist}$ between these two probability distributions was computed for every edge and HCP task considered in this study. This edgewise functional connectivity distance from resting-state can be seen as task-specific connectivity distance. That is, how far is the distribution of values in a specific task with respect to the resting-state FC baseline.    

\hl{The Jensen Shannon distance as defined above is thought for the general case where the two distributions P and Q come from different data samples (e.g. different subjects). This is not the most proper assessment in situations (like in the HCP data) where repeated measurements from same subjects are available, allowing for paired comparisons between FCs. To cover the necessity of that, we extended the concept of Jensen Shannon distance by proposing a "paired" version of $JS_{dist}$, namely $JS_{dist}^{paired}$, as follows. Similarly to $JS_{dist}$, for every edge in a functional connectome, we extract the corresponding individual values. Then we take the element-wise difference between the two vectors, where subjects are indexed in the same order. Such difference vector is then used to obtain distribution $P$ ($[-2,2]$ range, with uniform binning size of $0.1$). We then compare $P$ against a "null" distribution $Q$ which is probability $1$ at the bin including the zero value, and zero for all other bins. Such $Q$ distribution reflects the expectation of having no difference between repeated measurements. Deviations of $P$ with respect to $Q$ on functional edges reflect individual changes (in any direction and/or magnitude) between repeated measurements on subjects. Finally, we iterate this procedure for every edge and task to compute $JS_{dist}^{paired}$.
}

\subsection*{Centralized and distributed  processing in functional connectomes}
\label{cpdp}

The aforementioned procedure produced $374\times374$ (i.e. number of regions in the employed brain atlas) $JS_{dist}$ matrices per each task. Next, we sought to relate the proposed connectivity distance measure with changes in functional processing across functional networks (FNs). To do so, we first thresholded the $JS_{dist}$ matrices based on the $95$th percentile of the entire $JS_{dist}$ distribution of values across all tasks, to select only the most distant links from resting-state (see Fig. \ref{fig2}). Next, we quantify the amount of change in each of the $8$ functional FNs (see \nameref{functional data} section for details on the chosen FNs) by counting the number of edges that survived the threshold divided by the total number of edges present in each FN. We can then formalize changes in centralized processing (CP), for each functional network $k$, as:    

\begin{equation}
  CP^{k} = \frac{\sum\limits_{i,j \in WN^{k}} \widehat{JS_{dist}(i,j)}}{\sum\limits_{i,j \in WN^{k}} WN^{k}(i,j)}
\end{equation}  

Where $\widehat{JS_{dist}}$ is the binary version of the $JS_{dist}$ matrix thresholded at $95$ percentile ($1$ for surviving edges, $0$ elsewhere) for a specific task, and $WN^{k}$ is a binary matrix of the same size as $\widehat{JS_{dist}}$, with $1$ if an edge falls within functional network $k$, and $0$ elsewhere. 
Similarly, one can quantify changes in distributed processing (DP) as:

\begin{equation}
  DP^{kl} = \frac{\sum\limits_{i,j \in BN^{kl}}\widehat{JS_{dist}(i,j)}}{\sum\limits_{i,j \in BN^{kl}} BN^{kl}(i,j)}
\end{equation}  

Where now $BN^{kl}$ is a binary matrix of the same size as $\widehat{JS_{dist}}$, with $1$ if an edge falls between functional networks $k,l$, and $0$ elsewhere.

Hence, for each one of the $8$ functional networks considered here (see \nameref{functional data} for details), one can obtain one value of CP and $7$ values of DP (considering all pairwise FNs interactions), for a specific task. These values provide an estimate of the density of the most connectivity distant functional links across within and between FN connectivity. That is, the amount of local changes in distributed and centralized processing in each FN from baseline, defined as resting-state functional connectivity.       
\subsection*{\hl{Bandpass filter evaluation on centralized and distributed processing analysis}}
\hl{
In order to check if the different bandpass ranges applied for resting-state and task had an impact on centralized and distributed processing in FCs, we applied the same (liberal, i.e. $[0.001 Hz , 0.25 Hz]$) bandpass filter onto resting state data. We then evaluated changes in centralized and distributed processing after the new bandpass, and check the similarity with the ``standard'' bandpass results by computing the cosine similarity between the vectors defined by centralized and distributed processing coordinates. Cosine similarity is a measure of similarity between two non-zero vectors of an inner product space that measure the angle between them. Here, each vector represents a point in the state space defined by centralized and distributed processing.
}
\subsection*{Null models evaluation for connectivity distance analysis}
\label{surrogates}
To validate the $JS_{dist}$ results in functional connectomes, we tested the same approach on randomized counterparts (or ``surrogates'') of the original data. To do so, we employed the Amplitude Adjusted Fourier Transform (AAFT) surrogates method \cite{schreiber_surrogate_2000} to obtain data random surrogates. Starting from the 374 fMRI time series (one per brain region in the atlas, see also \nameref{brain atlas}) we generated AAFT fMRI time series surrogates as
proposed in \cite{schreiber_surrogate_2000}. This method aims to build surrogate time series that preserve the amplitude distribution and the power spectrum of the original data \cite{schreiber_surrogate_2000}. 

For each of the seven HCP tasks and resting-state, we computed 100 surrogate versions of the functional connectivity matrices, and then evaluated number of non-zero elements in $\widehat{JS_{dist}}_{\mbox{\tiny{surrogate}}}$ for each of the $100$ realizations. This provided null distributions (one per task) for the connectivity distance measure, that allowed us to test whether or not the results obtained on the original FCs were statistically significant.

\hl{Finally, in order to test the significance of differences between pairs of (highly-structured) correlation matrices, we also employed a permutation test that preserved intact the correlation structure of the FC matrices but randomly permuted the task/rest labels, computed the JS distance on the permuted data and then evaluated the number of JS edges per task surviving the $95$\% threshold based on the original data.}

\subsection*{Estimation of functional connectivity distance associations with structural connectomes}
\label{jsdsc}
Next, we sought to assess the role of structural connections in the connectivity distance of functional links across all seven tasks. In order to do so, we divided the group-averaged structural connectivity (SC) weights (see \nameref{structuraldata} for details on SC computation) into $5$ different percentile intervals \{$0-20$; $20-40$; $40-60$; $60-80$; $80-100$\}. We then counted the average number of most distant edges (i.e. the non-zero elements of $\widehat{JS_{dist}}$) falling in each of the $5$ percentile intervals, for each of the seven HCP task. This provides an estimate on the relationship between structural connections and connectivity distance and whether it depends on the specific task being performed. We also tested whether centralized and distributed processing depend on the ``nestedness'' or ``hiddenness'' of the structural pathways, as measured by search information \cite{goni_resting-brain_2014,trusina2005communication,rosvall2005searchability}. Search information \hl{(SI)} quantifies the accessibility or hiddenness of the shortest path between a source node and a target node within the network by measuring the amount of knowledge or information in bits needed to access the path \cite{goni_resting-brain_2014,trusina2005communication,rosvall2005searchability,wirsich2016whole} The more nested the shortest path between two brain regions $ij$, the higher its SI value; conversely, the less hidden or integrated the path, the lower its the SI value.
Similarly to the experiment performed on SC weights, we again divided the group-average SI range of values into $5$ different percentile intervals: \{$0-20$; $20-40$; $40-60$; $60-80$; $80-100$\}. Finally, we counted the average number of most distant edges (i.e. the non-zero elements of $\widehat{JS_{dist}}$) falling in each of the 5 SI percentile intervals, for each of the seven HCP task. This provides an estimate on the relationship between structural ``hiddenness'' and connectivity distance and its associations with the specific task being performed. For both SC weights and SI, the significance of the associations with centralized and distributed processing was assessed through one-way analysis of variance (ANOVA \cite{hogg1987engineering}, Matlab command \textit{anova1}), with ``observations" being centralized and distributed processing values for the $7$ tasks, and ``groups" being the $5$ percentile intervals described above.

\section*{Results}
\label{results}

\begin{figure}[!ht] 
\centerline{\includegraphics[width=\textwidth]{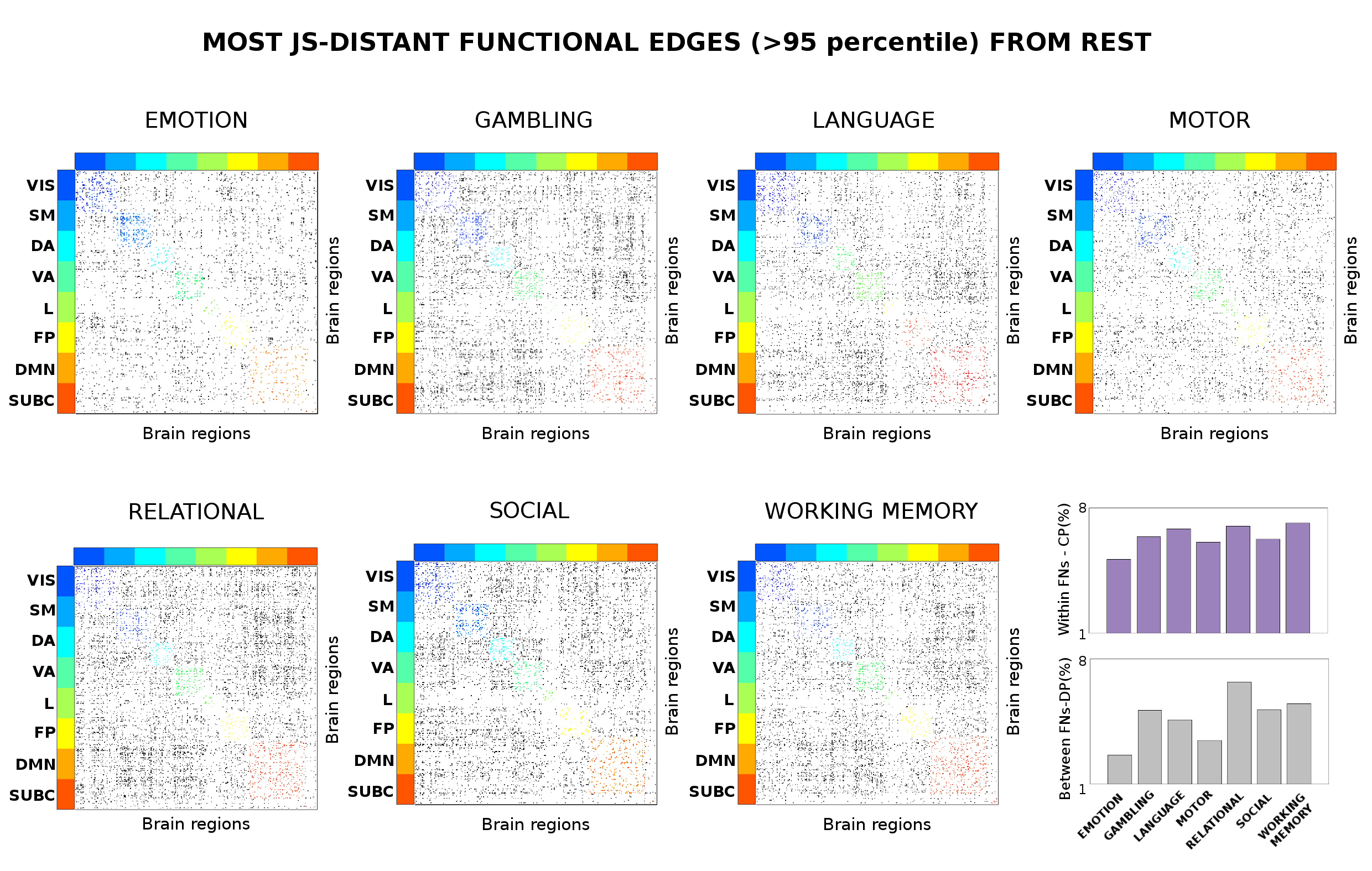}}
\caption{\textbf{Connectivity distance across different tasks.} Evaluation of the most distant functional links (in terms of Jensen-Shannon (JS) distance, see Methods) across $7$ different task sessions. The JS matrices were thresholded at the $95$\% of the distribution of JS values across the seven tasks. The JS matrices then ordered by $7$ functional networks (FNs, \cite{yeo_organization_2011}): visual (VIS), somato-motor (SM), dorsal attention (DA), ventral attention (VA), limbic (L), frontoparietal (FP), default mode network (DMN). An eight subcortical network (SUBC) was added for completeness. The edges surviving the threshold corresponding to within-FN connections color-coded accordingly. Edges corresponding to between-FN connections are depicted in grayscale. Note how the connectivity distance depends on the task: in some cases within-FNs connectivity are more recruited (i.e., for the Emotion task), in other between-FNs connections are the most distant (i.e., Relational task). The bottom-right bar plots depict the average percentage of within-FNs most distant edges, i.e. centralized processing (CP, violet bars) and the average percentage of between-FNs edges, i.e. distributed processing (DP, grey bars) across the different tasks.}  
\label{fig2}
\end{figure}

The dataset used for this study consisted of functional data from the 100 unrelated subjects in the Q3 release of the HCP \cite{van_essen_human_2012,van_essen_wu-minn_2013}. We defined the ``connectivity distance" between task FC links and resting-state FC links as the edgewise Jensen Shannon distance ($JS_{dist}$) between resting-state FCs and task FCs (see also scheme at Fig. \ref{fig1}). This metric quantifies the connectivity distance of a functional link recruited in a task with respect to its correspondent ``usage'' in resting-state. For each of the $7$ HCP tasks (see \nameref{methods} for details), we computed the corresponding $JS_{dist}$ and \hl{{$JS_{dist}^{paired}$}} matrices, and extracted the most connectivity distant edges ($\geq 95$ percentile distribution of $JS_{dist}$ and \hl{{$JS_{dist}^{paired}$}} values across all tasks. \hl{Fig. \mbox{\ref{fig2}} shows the results corresponding to {$JS_{dist}$} (unpaired, all different subjects for resting-state and task FCs), and Fig. S1 summarizes the results for {$JS_{dist}^{paired}$} (paired, same subjects for resting-state and task FCs). Please see also \mbox{\nameref{jsd}} for details}.

Notably, the results obtained are significantly different from the same analyses performed on 100 realizations of surrogate data built from the fMRI time series considered in this study (Fig. S2 and Table S1) , see \nameref{surrogates} for details). Furthermore, with the only exception of MOTOR versus EMOTION for absolute frame displacement, no significant differences were observed in frame-wise displacement estimates ($p<0.01$, double-sided t-test between task pairs). This included absolute frame displacement (root mean squared, HCP filename Abs\_RMS) and relative frame displacement (root mean squared, HCP filename Rel\_RMS). These findings suggest that head motion is not biasing rest to task JS distances depicted in Figure 2 and Figure S1.

\begin{figure}[!ht] 
\centerline{\includegraphics[width=\textwidth]{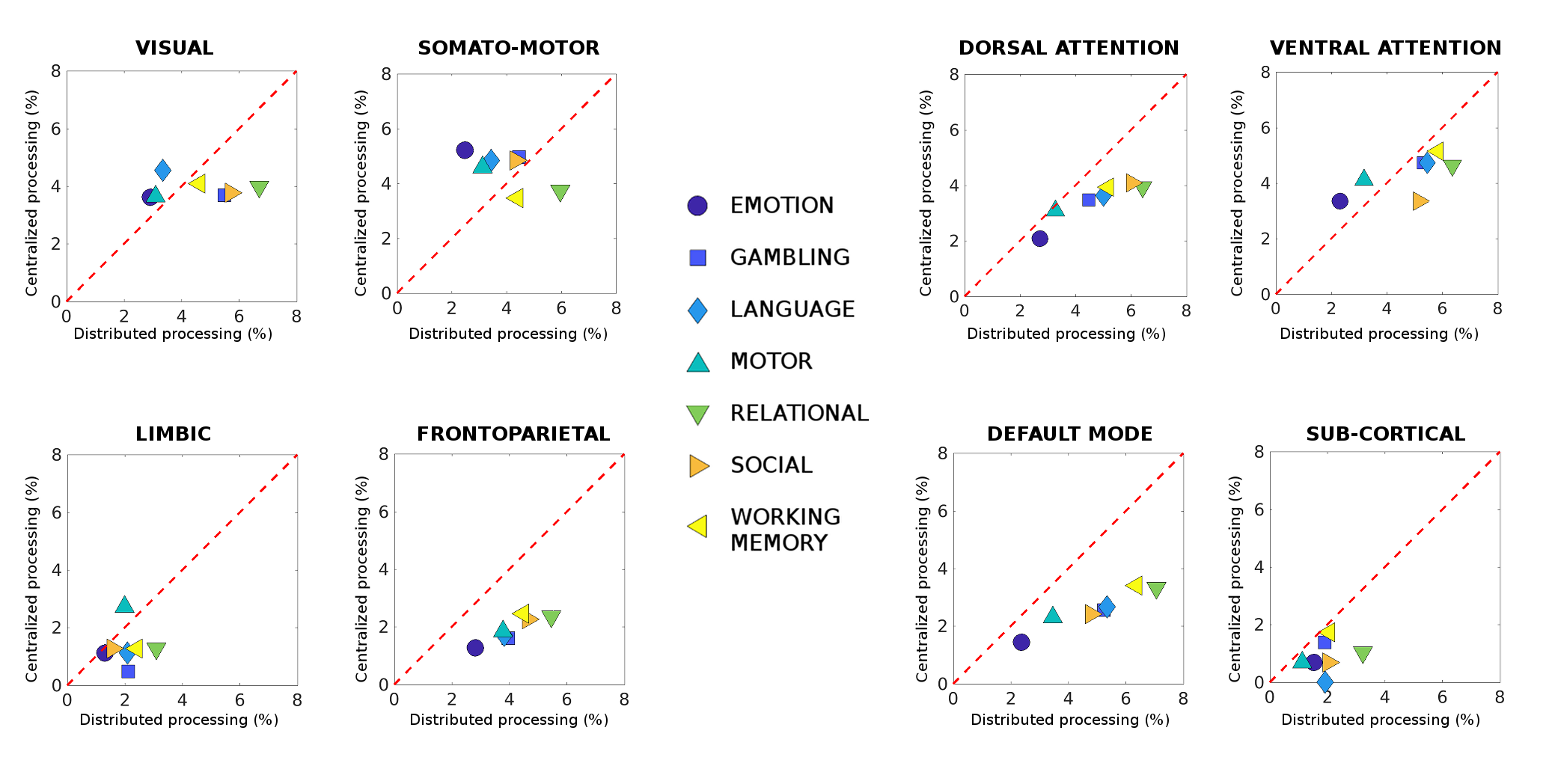}}
\caption{\textbf{Centralized and distributed task processing in functional connectomes.} Each plot shows differences in centralized versus distributed processing (see Methods) for each of the seven functional networks (FNs,visual, somato-motor, dorsal and ventral attention, limbic, frontoparietal, DMN, \cite{yeo_organization_2011}) and sub-cortical network, for all the seven different HCP tasks. The difference in centralized processing with respect to resting state was defined as the number of most Jensen-Shannon (JS) distant edges within-FN divided by the total number of edges in the FN (reported as percentage). Similarly, deviations form distributed processing in resting-state were defined as the number of most JS-distant edges between FN divided by the total number of between FN connections. Note how FP and DMN networks deviate from rest mainly in the amount of distributed processing, i.e. between-FNs connectivity.}
\label{fig3}
\end{figure}

Interestingly, the level of distance from resting state seems to be associated to the specific task (Fig. \ref{fig2}). For some task, the within-functional network links are more distant, i.e. more involved (e.g., for the Emotion and Motor tasks), in other the between-FNs connections are the most distant ones (i.e., Relational or Working memory tasks). The dichotomy between intra-network (i.e. centralized) and inter-network (i.e. distributed) distance led us to quantify the changes in centralized and distributed processing in task FCs (Fig. \ref{fig3}, see also the \nameref{cpdp} section). 

Note how, for three functional networks, i.e. dorsal, frontoparietal and default mode, there is a clear demarcation between centralized and distributed processing, for all the seven tasks evaluated (Fig. \ref{fig3}). \hl{This indicates that the functional connections between these networks get more distant from rest when they are recruited in a task}. Furthermore, with the exception of limbic and subcortical networks, where little difference in centralized and distributed processing can be observed (Fig. \ref{fig3}), in all the other FNs (i.e., visual, somato-motor and ventral attention) there is balance between intra and inter-network processing. This trade-off seems to depend on the task at hand (slightly more centralized in some, more distributed in others, Fig. \ref{fig3}). \hl{Note that these different distributions of centralized and distributed processing across tasks are not related to the different bandpass applied for rest and task data, since very similar results were obtained when the same (liberal, i.e. $[0.001 Hz , 0.25 Hz]$) bandpass parameters were applied onto resting-state data (see Fig. S3 in supportive information).}  

\begin{figure}[!ht] 
\centerline{\includegraphics[width=0.95\textwidth]{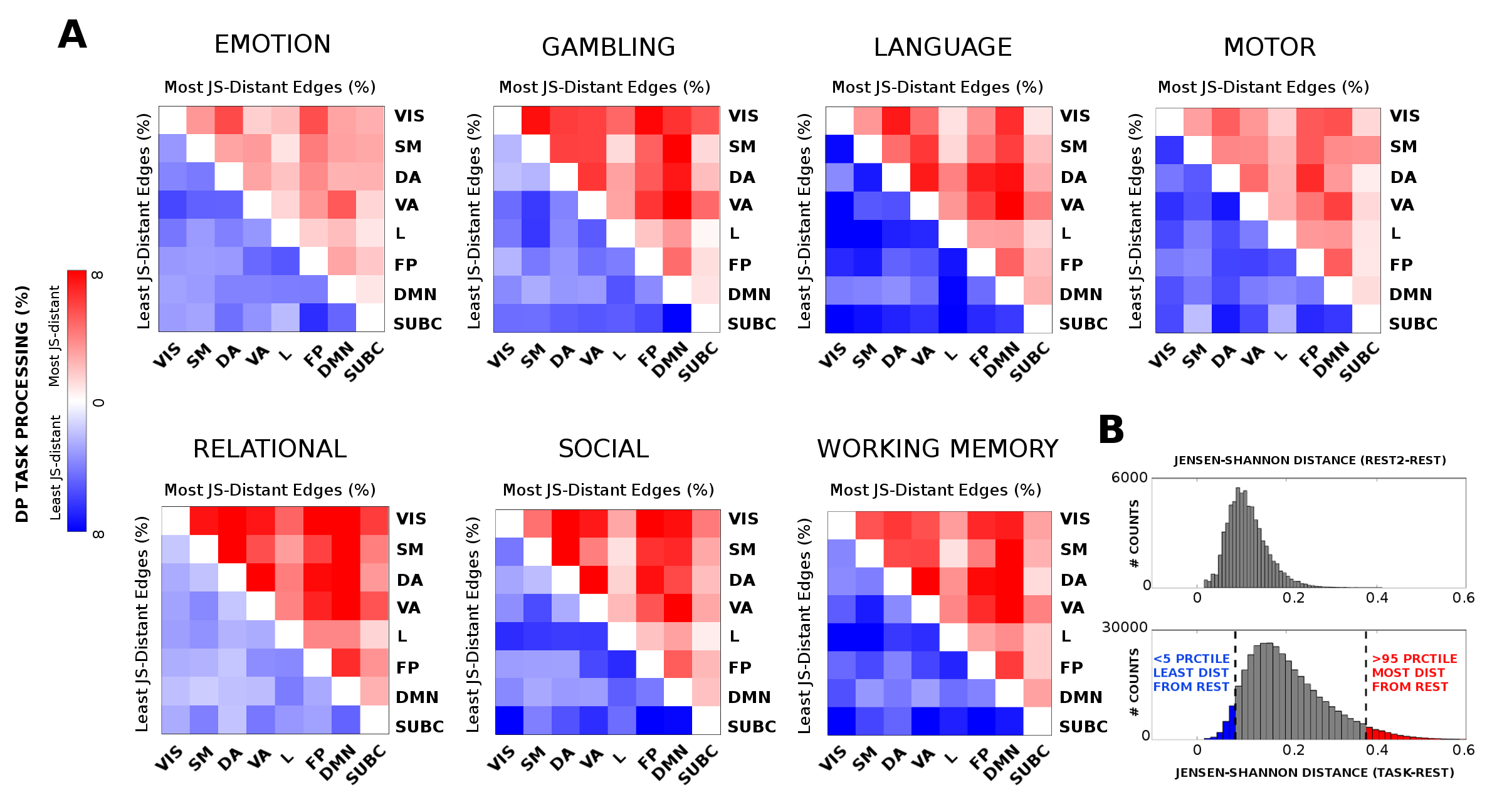}}
\caption{\textbf{\hl{Least and most distant edges per functional network across tasks.}} \textbf{A)} \hl{ Heat maps, for all seven fMRI tasks evaluated, showing the most (red, upper triangular) and least (blue, lower triangular) distributed processing (DP) values between pairs of functional networks with respect to REST. \textbf{B)} Top: distribution of JS distance values when comparing REST2 session to the baseline REST session. Bottom: distribution of Jensen-Shannon distance values across the seven tasks evaluated. The tails of the histogram are highlighted in red (least distant edges, $<$5th percentile) and blue ( most distant edges, $>$ 95th percentile).} 
}
\label{fig4}
\end{figure}

\hl{When looking at JS-distance differences between functional networks across tasks, it may be observed task specific patterns (see Figure \mbox{\ref{fig4}A)}. For instance, Relational task exhibit a whole-brain tendency to get more distant from rest (i.e. higher general distributional processing); Emotion or Motor tasks are among the least distant from rest; whereas some other tasks (e.g. Language or Social) display a more specialized distributed processing across functional networks combining most and least distant functional edges. The histogram of the distribution of JS-distance values across all tasks gives more insights on the cognitive distance task-rest: there is a general tendency to be different from REST. However, some edges ($<5$\%, Figure \mbox{\ref{fig4}B)}, blue bars) stay almost unchanged with respect to REST configuration; some others ($>95$\%, Figure \mbox{\ref{fig4}B)}, red bars) switch to more distant values, allowing for the cognitive reconfiguration of the system. Note that the JS-distance does not depend of the baseline (i.e. REST FCs) magnitude of the correlations, as the average correlation between JS-values and median REST FC correlation values across tasks is $-0.006 \pm 0.026$. Importantly, no single edge survived to the 95th percentile threshold after permutation testing of the TASK-REST labels (see Methods for details) (Fig. S4). Analogously, less than 1\% of REST2-REST edges survived to that same threshold (Fig. 4B). Overall, these analyses indicate that the conservative threshold chosen will depict actual task-rest FC changes.}

\begin{figure}[!ht] 
\centerline{\includegraphics[width=0.9\textwidth]{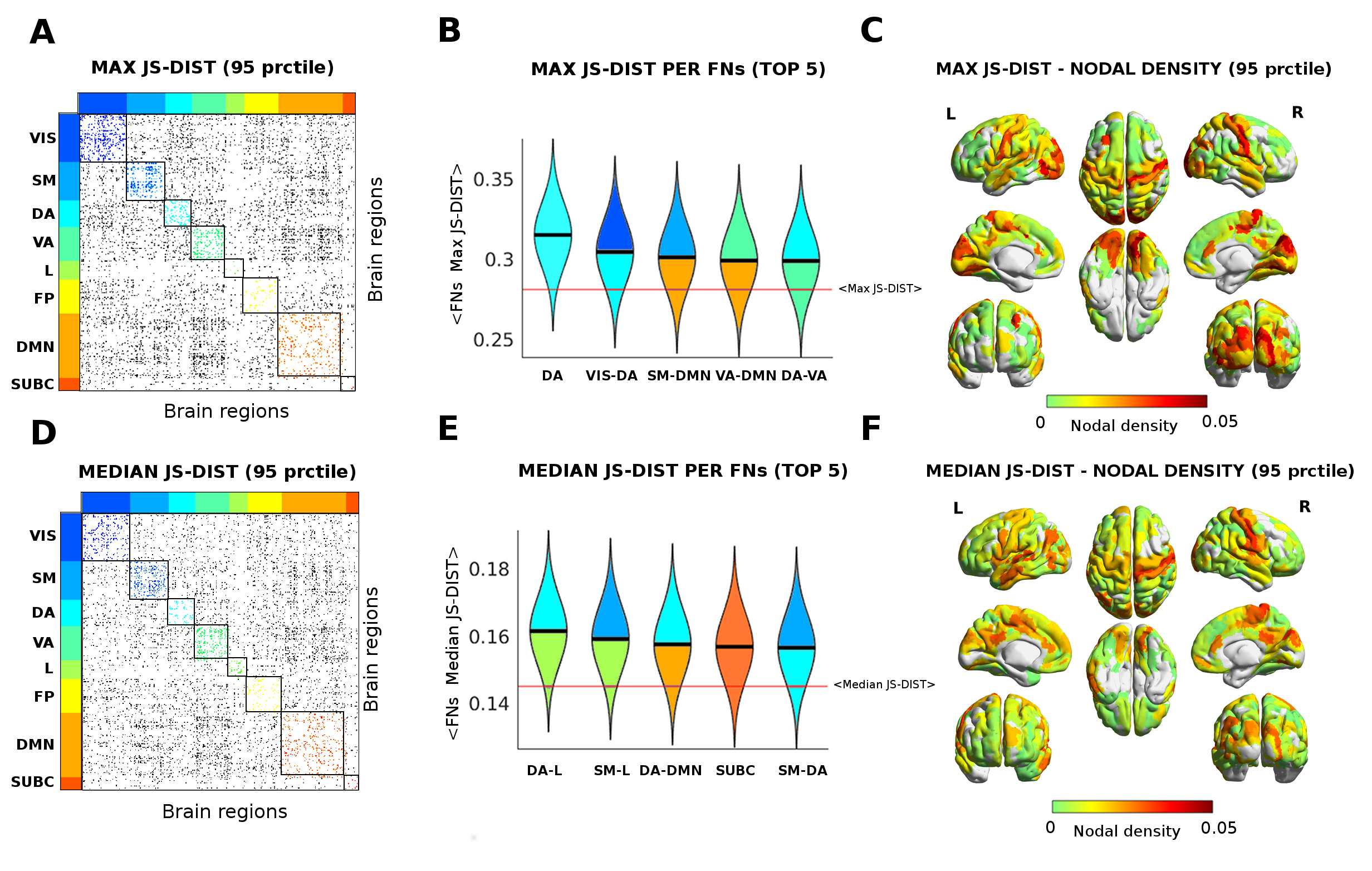}}
\caption{\textbf{\hl{Functional reconfiguration via Jensen-Shannon distance.}} \textbf{A-D)} \hl{Edgewise max (A) and median (D) Jensen Shannon distance across all tasks (thresholded by 95th percentiles for max and for median). The colored dots depict JS values within FNs; gray dots indicate significant JS-distant edges between FNs. \textbf{B-E)} Violin plot of edgewise JS-distance (max and median) for the top 5 FNs and FNs interactions. Within FNs are color-coded accordingly (as in A-D), while between FNs are color-coded using the colors of the two FNs involved. Solid black lines depict median values of each distribution;  solid red line indicates the whole-brain median value of max and median distributions.\textbf{C-F)} Brain render of max and median JS distances as nodal density per region. The strength per brain region computed as sum of JS-distance (max and median) for functional edges above the 95 percentile threshold divided by the total number of brain regions.}   
}
\label{fig5}
\end{figure}

\hl{The results in Figure \mbox{\ref{fig4}}, display the extent to which JS distance is sensitive to different levels of functional reconfiguration} \cite{schultz2016higher,krienen2014reconfigurable,shine2016dynamics} \hl{for different functional networks across different tasks. Hence we decided to explore further on this, and evaluated the edgewise maximum and median JS-distance across all tasks. This would provide an overall summary of the main edges and functional networks involved in the cognitive switch across the seven tasks (see Figure \mbox{\ref{fig5}}). Notice that median and max give two different ``flavors'' of brain network reconfiguration: the median tells us, how far, on average, brain regions (and functional networks) change when subjects perform tasks. The max tells us what are the most distant values, hence the highest achieved functional reconfiguration for a brain region across the seven different tasks under study.}

In order to determine whether changes in task processing are related to the underlying structural connectivity, we first evaluated the relationship between connectivity distance in each task and structural connectivity weights (Fig. \ref{fig4}, A1-B1). Interestingly, a significant trend arises for all task between centralized processing and number of tracts \hl{(one-way ANOVA $F=163.39$, $df =4$, $p=6.62 \cdot 10^{-20}$}, Fig. \ref{fig4}, A1). That is, the more structurally connected two regions are within a functional network, \hl{the higher the number of centralized processing edges recruited in a task}. Notice how this trend is general and independent from the task, albeit the magnitude of this linear association between structure and cognitive depends on the task at hand (Fig. \ref{fig4}, A1).
On the other hand, when looking at changes in distributed processing, i.e. for edges involved in between-functional network connectivity, no significant associations with structural connectivity were observed (\hl{one-way ANOVA $F = 1.11$, $df = 4$, $p=0.37$}, Fig. \ref{fig4}, B1).  

\begin{figure}[!ht] 
\centerline{\includegraphics[width=1\textwidth]{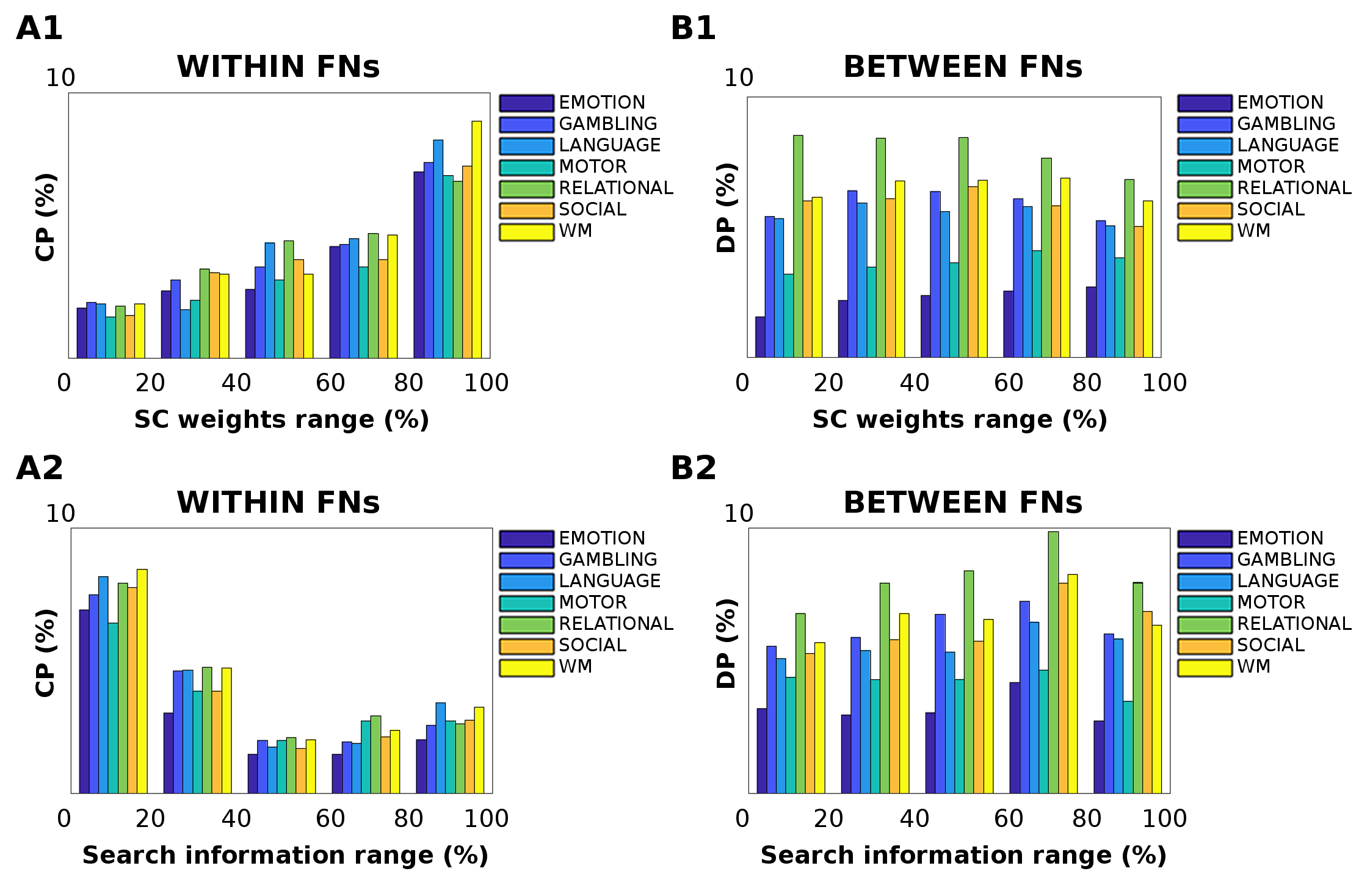}}
\caption{\textbf{Effect of structural pathways on centralized and distributed processing changes.} \textbf{A1-B1)} \hl{The relationship between the anatomical connections and Jensen-Shannon distance was evaluated across the seven different tasks}. The bar plots shows the percentage of centralized processing (CP) within functional networks (FNs, A1) and distributed processing (DP) between FNs (B1), per 5 different percentile range of structural connectivity weights: $0-20$, $20-40$, $40-60$, $60-80$ and $80-100$. The percentile range was extracted from the group-averaged structural connectome. Note how, for within-FNs connections (A1), the change in centralized processing significantly correlates with the strength of structural connections across all tasks (\hl{one-way ANOVA $F=163.39$, $df =4$, $p=6.62 \cdot 10^{-20}$}); conversely, the underlying structural connectivity does not play a major role in distributed processing changes \hl{(one-way ANOVA $F = 1.11$, $df = 4$, $p=0.37$).} \textbf{A2-B2)} The effect of structural path accessibility (as measured by search information, see Methods) on centralized and distributed processing was tested across the seven different tasks, per 5 different percentile intervals of search information: 0-20, 20-40, 40-60, 60-80 and 80-100. The percentile range was extracted from the group-averaged search information matrix. Notably, change in centralized processing (A2) are significantly associated to low values of search information (one-way ANOVA \hl{one-way ANOVA $F = 131.75$, $df = 4$, $p=1.41 \cdot 10^{-18}$}); \hl{conversely, no significant association between SI and distributed processing changes was found (one-way ANOVA $F = 1.85$, $df = 4$, $p=0.14$)}}
\label{fig6}
\end{figure}

We then dug deeper into the relationship between task processing and structural connectome by evaluating the level of hiddenness or accessibility of a structural path, as measured by search information \cite{goni_resting-brain_2014} (see also \nameref{jsdsc} for details), and testing its association with changes in cognitive task processing (Fig. \ref{fig4}A2-B2). Interestingly, the hiddenness of structural paths appears to be inversely related to changes in centralized processing (\hl{one-way ANOVA $F = 131.75$, $df = 4$, $p=1.41 \cdot 10^{-18}$}, Fig. \ref{fig4}, A2). That is, the more ``isolated'' the structural pathway between two brain regions within a functional network, the higher will be its recruitment in a task. The more nested or integrated the path, the less distant the centralized processing with respect to resting state (Fig. \ref{fig4}, A2). No significant associations were found when looking at changes in distributed processing versus search information range of values (\hl{one-way ANOVA $F = 1.85$, $df = 4$, $p=0.14$}, see Fig. \ref{fig4}, B2).


\section*{Discussion}

Cognitive brain network mapping \cite{cole_cognitive_2007,cole_multi-task_2013,krienen2014reconfigurable,finn2017can}, or the analysis of brain network features underlying task performance and cognitive control \cite{cole_intrinsic_2014,tavor_task-free_2016,finn2017can,gratton2016evidence,Khambhati2018}, is a recent and exciting new line of investigation in brain connectomics. While the general intrinsic common architecture between resting state and task-based functional patterns has been explored \cite{cole_intrinsic_2014}, still very little is known about task connectivity distances and their associations to information processing \cite{ito_cognitive_2017,cole_cognitive_2007,cole2016activity}. Furthermore, an even more intricate question relates to the relationship between the different task-based FC scenarios and the underlying structural connectivity \cite{fukushima2018structure,hermundstad2013structural,hermundstad2014structurally,mivsic2016network,amico_mapping_2017-1}.  

Here we addressed these questions by proposing a novel methodology in neuroscience, rooted to the concept of Jensen-Shannon divergence \cite{de_domenico_structural_2015,briet_properties_2009}, to measure task-based pairwise functional distance with respect to the ``cognitive baseline'' defined by resting-state FCs (Fig. \ref{fig1}). This framework may also be seen from a multilayer perspective, with the ground layer being resting-state FCs and top layers defined by the multiple task-based different connectivity scenarios. The distance defined here can be thought as inter-layer coupling, or as the amount of cognitive processing necessary to make the ``cognitive switch'' from the resting-state ground layer to the top task-based functional layers. \hl{The JS divergence has several advantages: it is a non-parametric test, does not assume any form of distribution and allows for quantifying fine-grained changes between two distributions. As shown in Figure S5, there were a large number of functional edges for which REST or TASK FC distributions did not pass a normality test.}  

The work presented here complements the aforementioned recent studies on cognitive mapping, where the resting state scaffolding was usually used to infer or also predict task changes in connectivity \cite{cole_intrinsic_2014,tavor_task-free_2016,ito_cognitive_2017,cole2016activity}. Here we evaluate and investigate the pairwise distance task-rest, and use it to map specific changes dictated by the task at hand. This adds up to previous studies in that it improves our understanding of how edge specific is the cognitive switch, and its level of recruitment (in terms of ``connectivity distance''), as well as in terms of centralized and distributed processing changes in functional networks (Fig. \ref{fig2}).   

We exploited this new concept of connectivity distance to infer about the level of recruitment of an edge or of a functional network (Fig. \ref{fig2}). Notably, the connectivity distant patterns present in the original data were significantly different from the ones obtained by surrogate data built from the original fMRI time series (Fig. S2 and Table S1) or \hl{from the ones obtained by randomly shuffling rest-task FCs}. Indeed, the more distant a functional network in a specific task, the more different its recruitment with respect to resting-state. Hence, the more changes in cognitive information processing that functional subsystem will undergo. This intuition led us to explore the concept of centralized and distributed processing in large scale functional networks, that we defined as the difference between intra (i.e. centralized)-and inter (i.e. distributed) network connectivity (Fig. \ref{fig3}). Interestingly, \hl{three functional networks (dorsal attention, frontoparietal and DMN)} showed major changes in distributed processing and very minor changes in centralized processing, for all the seven tasks evaluated with respect to resting state (Fig. \ref{fig3}). This is in line with recent findings showing that frontoparietal and attentional areas appear to be the more flexible for cognitive control and task performance \cite{cole_cognitive_2007,cole_multi-task_2013,shine2016dynamics,krienen2014reconfigurable}. 

The fact that these networks and the DMN, which is well-known to play a major role in resting-state \cite{raichle_brains_2015,raichle_default_2001,greicius_functional_2003}, change mainly in terms of inter-communication when transitioning to task, is also noteworthy. This finding goes along with the concept of integration of information between neural subsystems \cite{tononi_measure_1994} and also with our recent findings on the association between FP-DMN dis-connectivity and degradation in arousal and levels of consciousness \cite{amico_mapping_2017}. Possibly, the more demanding the task, the more the cross-talk between FP, DMN, \hl{attentional networks} and rest of the brain might be needed to achieve the proper amount of cognitive processing \hl{or ``brain network reconfiguration''} \mbox{\cite{schultz2016higher,krienen2014reconfigurable}}. 

\hl{We further investigated on the concept of brain network reconfiguration across tasks, by evaluating most and least distant functional edges between functional networks (Figure \mbox{\ref{fig4}}). Interestingly, some tasks seem to require extremely distant interactions between FNs (e.g. Relational), some other tasks require specific subsets of FNs interactions (e.g. Language, Working Memory). The investigation of the max and median nodal $JS_{dist}$ centrality (Figure \mbox{\ref{fig5}}) across tasks added more information on the complex scenario depicted in Figure \mbox{\ref{fig4}}. Dorsal and occipital regions seem to be the ones that are generally more distant from rest when engaged in a task (Figure \mbox{\ref{fig5}}, A-C). On the other hand, somato-motor and dorsal areas seem to be the ones that achieve the largest reconfiguration (i.e. maximal $JS_{dist}$ from resting-state) across the seven tasks evaluated (Figure \mbox{\ref{fig5}}, D-F).}

\hl{Taken together, these findings suggest that the cognitive ``switch'' between resting-state and task states is more than a general shift in terms of functional links, but rather a complex interplay between maximally distant and minimally distant functional connections (Figure \mbox{\ref{fig4}}, Figure \mbox{\ref{fig5}}). This is in line with recent studies investigating the complex reconfiguration of brain networks during tasks} \cite{bassett2013task,shine2016dynamics,krienen2014reconfigurable,Khambhati2018}.

Another major question relates to how these changes in cognitive processing are shaped or determined by the underlying structural architecture of a human brain. Very few studies so far have tried to elucidate the relationship between cognitive changes and axonal pathways, either for localized cortical subsystems (e.g. fusiform gyrus \cite{saygin2012anatomical}) or for a specific task (e.g. visual stimuli \cite{osher2015structural}), \hl{or at the whole-brain level \mbox{\cite{hermundstad2013structural,hermundstad2014structurally,mivsic2016network}}}. In a recent work we tackled this problem from in a whole-brain network fashion, by means of ICA-based approach to extract the main ``hybrid'' functional-structural connectivity features sensitive to cognitive changes across seven different tasks \cite{amico_mapping_2017-1}. 

Here we took this investigation one step further by assessing functional connectivity distance associations with respect to the underlying structural connectivity weights (Fig. \ref{fig4}). Interestingly, for changes in centralized processing, the relationship with structural connectivity is linear (Fig. \ref{fig4}A1). That is, when the cognitive processing involves mainly within-network connectivity, the higher the fiber strength between two regions, the more distant they will be. Nonetheless, this relationship is not present when looking at distributed processing link to link effects (Fig. \ref{fig4}B1). This might imply that the between network connectivity links can play a key role in the creation of more complex cognitive regimes, \hl{in line with previous findings \mbox{\cite{fukushima2018structure,hermundstad2013structural,hermundstad2014structurally,mivsic2016network}}} . The cross-talk between functional networks might bring the brain network up to a more integrated level, allowing for a more dynamic and distributed cognitive processing, that ultimately deviates far from the static underlying boundaries given by the structural fiber tracts.

To test this hypothesis, we evaluated the relationship between changes in distributed and centralized processing and structural path ``hiddenness'' or accessibility, as measured by search information (\cite{goni_resting-brain_2014}, see also \nameref{jsdsc}). Notably, when looking at centralized processing deviations from rest, these two quantities appear to be inversely related (Fig. \ref{fig4}A2,B2). The less integrated the path between two regions within a functional network, the higher the value in centralized processing, the more integrated the structural pathways, the less centralized activity (Fig. \ref{fig4}A2). 

These findings corroborate the hypothesis on the integration segregation in the human brain \cite{tononi_measure_1994}. It is also in line with our findings on the importance of cross-talking between functional networks for task changes (\cite{amico_mapping_2017-1}), which can be summarized as follows: for a human brain to make a cognitive switch, a delicate interplay between centralized and distributed processing is necessary. The centralized activity within functional subsystems is shaped by brain structure. Moreover, the more isolated the shortest path connecting two centralized brain regions, the higher the level of task processing (Fig. \ref{fig4}). However, in order to achieve proper cognitive complexity for the task at hand, an appropriate level of distributed processing and subsequent integration between these subsystems is needed: the level of cross-talking and structural integration will depend on the specific task at hand (Fig. \ref{fig2}, Fig. \ref{fig4}) and on the functional subnetwork involved (Fig. \ref{fig3}), with no significant function-structure associations (Fig. \ref{fig4}).        
      
This study has several limitations. \hl{The framework presented here is based on the quantification of distances between estimations of functional connectivity data. The accuracy and representativity of the estimate JS-distance will be subject to the quantity and quality of the fMRI data as well as the processing steps. Further studies should explore how different aspects of the data and subsequent processing may have an impact in the JS-distance quantifications, including number of subjects, duration of the REST and TASK fMRI sessions, spatio-temporal limitations on fMRI data, and motion regressors included, among others. Analogously, our findings associating SC properties (weights and search-information on SC shortest-paths) with JS-distance may also be, to some extent, sensitive to diffusion tractography user-defined free parameters.} The effect of different brain atlases (here we used the one proposed by Glasser et al. \cite{glasser_multi-modal_2016}) and functional network organization (here we used the one proposed by Yeo et al. \cite{yeo_organization_2011}) on the centralized and processing changes should be explored. Because of the way it is defined, the measure does not allow for individualized connectivity distance patterns. However, future studies could explore to what extent individual FCs can be predicted by group-wise changes in centralized and distributed processing, or use the information on the most distant link at the group level to select the most meaningful pairwise connectivity for the task at hand. \hl{Another one potentially promising application of this methodology may be related to measure JS divergence at the single-subject level with dynamic functional connectivity. In this case, the different FCs would be given by the FC snapshots obtained from sliding the FC window along the same subject acquisition during resting-state and a specific task (as opposed to the slices being subjects as depicted in Figure {\ref{fig1}} of this paper).} 

\hl{It is also important to remark the importance of introducing a well-defined distance metric (i.e. $JS_{dist}$) in the functional connectome domain, as we propose in this work. This can open several promising new lines of research in brain connectomics in terms of topological analysis of the connectivity domain, and of the definition of metrizable spaces where to map cognitive changes in brain functional networks.} Finally, the simplicity of the methodology allows for its applicability in the clinical domain, where it could be employed to measure connectivity distance between ``healthy'' and ``diseased'' populations (e.g. Alzheimer, schizophrenia, coma etc.), or to determine task-rest distance in situations where the cognitive switch is damaged or disrupted (e.g. in autism or similar neurological disorders).    

In conclusion, we have reported a new methodology that aims at capturing the functional differences between different tasks when compared to resting state. The methodology based on the Jensen-Shannon distance is promising, and has been proved to discern between centralized and distributed activity across brain areas for different tasks. These results pave the way to the usage of this framework in other experiments, and to the development of a new information--theoretical framework for the analysis of functional and structural connectomes.


\section*{Acknowledgments}
Data were provided [in part] by the Human Connectome Project, WU-Minn Consortium (Principal Investigators: David Van Essen and Kamil Ugurbil; 1U54MH091657) funded by the 16 NIH Institutes and Centers that support the NIH Blueprint for Neuroscience Research; and by the McDonnell Center for Systems Neuroscience at Washington University. JG acknowledges financial support from NIH R01EB022574 and NIH R01MH108467 and the Indiana Clinical and Translational Sciences Institute (Grant Number UL1TR001108) from the National Institutes of Health, National Center for Advancing Translational Sciences, Clinical and Translational Sciences Award, Purdue Discovery Park Integrative Data Science Initiative. AA acknowledges financial support from Spanish MINECO (grant FIS2015-71582-C2-1), Generalitat de Catalunya ICREA Academia, and the James S. McDonnell Foundation (grant 220020325).

\section*{Author Contributions} 
E.A, A.A. and J.G conceptualized the study, designed the framework, interpreted the results and wrote the manuscript. E.A. and J.G. processed the MRI data and performed the analyses. 

\section*{Code availability}
The code used for computing Jensen-Shannon distance in functional connectomes will be made available on the CONNplexity lab website (\url{https://engineering.purdue.edu/ConnplexityLab}).


\bibliography{amico_et_al_TaskProc_main_SI_arXiv_v3}

\bibliographystyle{plain}

\newpage
\setcounter{figure}{0}
\renewcommand{\thefigure}{S\arabic{figure}} 

\section*{Supportive Information}

\begin{figure}[!ht] 
\centerline{\includegraphics[width=1\textwidth]{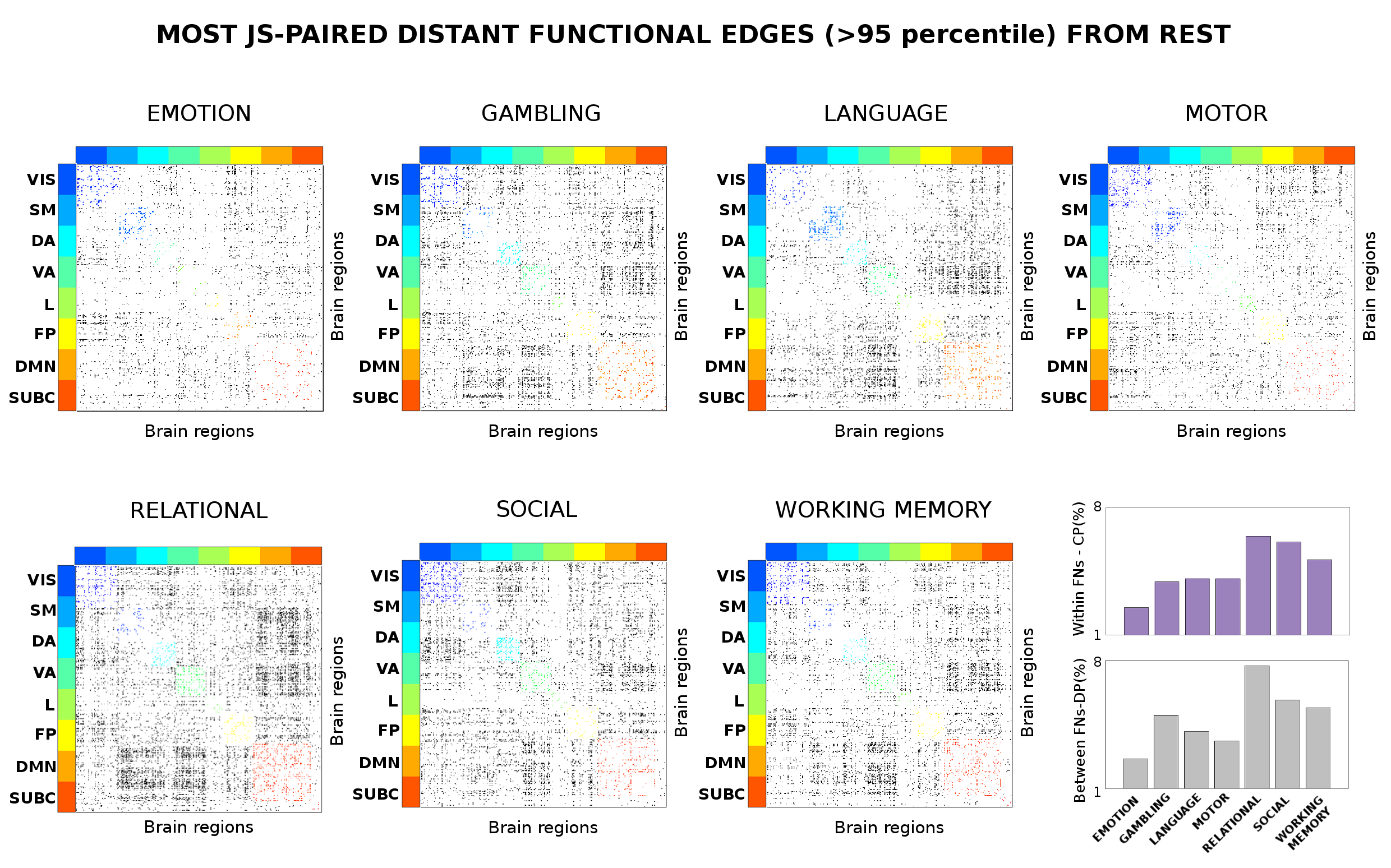}}
\caption{\textbf{Connectivity distance across different tasks (``paired'' Jensen-Shannon distance)} Results for Lilliefors test, edgewise for each task. Edges that did not pass the normality test (either at rest or on the assessed task) corresponding to within-FN are color-coded accordingly. Edges that did not pass the normality test corresponding to between-FN are depicted in grayscale. The percentage of non-normal edges per task is reported on top of each matrix. The bottom-right bar plots depict the average percentage of within-FNs non-normal edges (violet bars) and the average percentage of between-FNs non-normal edges (grey bars) for each task.
}
\label{figS1}
\end{figure} 

\newpage

\begin{figure}[!ht] 
\centerline{\includegraphics[width=1\textwidth]{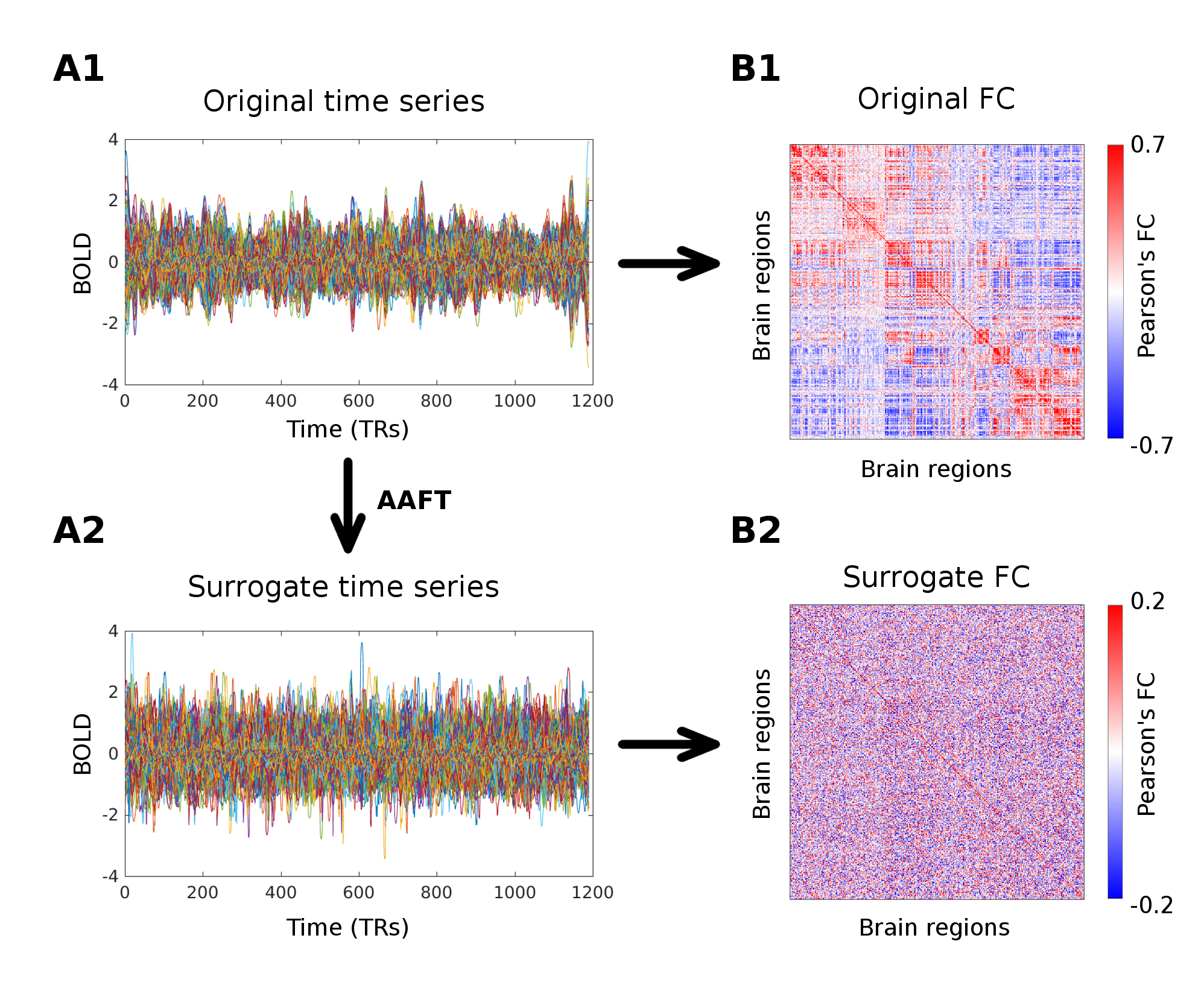}}
\caption{\textbf{Scheme of the functional connectome randomization procedure.} In order to validate the $JS_dist$ results in functional connectomes (FC), we used ``surrogates'' of the original data. We here show an example of this procedure for one subject, resting-state. The original BOLD 374 (i.e. one per brain region) time series (A1), from which the FC was originated (B1), were randomized by means of the Amplitude Adjusted Fourier Transform (AAFT) surrogates method (Schreiber and Schmitz, 2000), see section \ref{surrogates} for details). These randomized time-series (A2) were then used for the construction of the surrogate FC of the subject (B2).
}
\label{figS2}
\end{figure}
\newpage

\begin{figure}[!ht] 
\centerline{\includegraphics[width=1\textwidth]{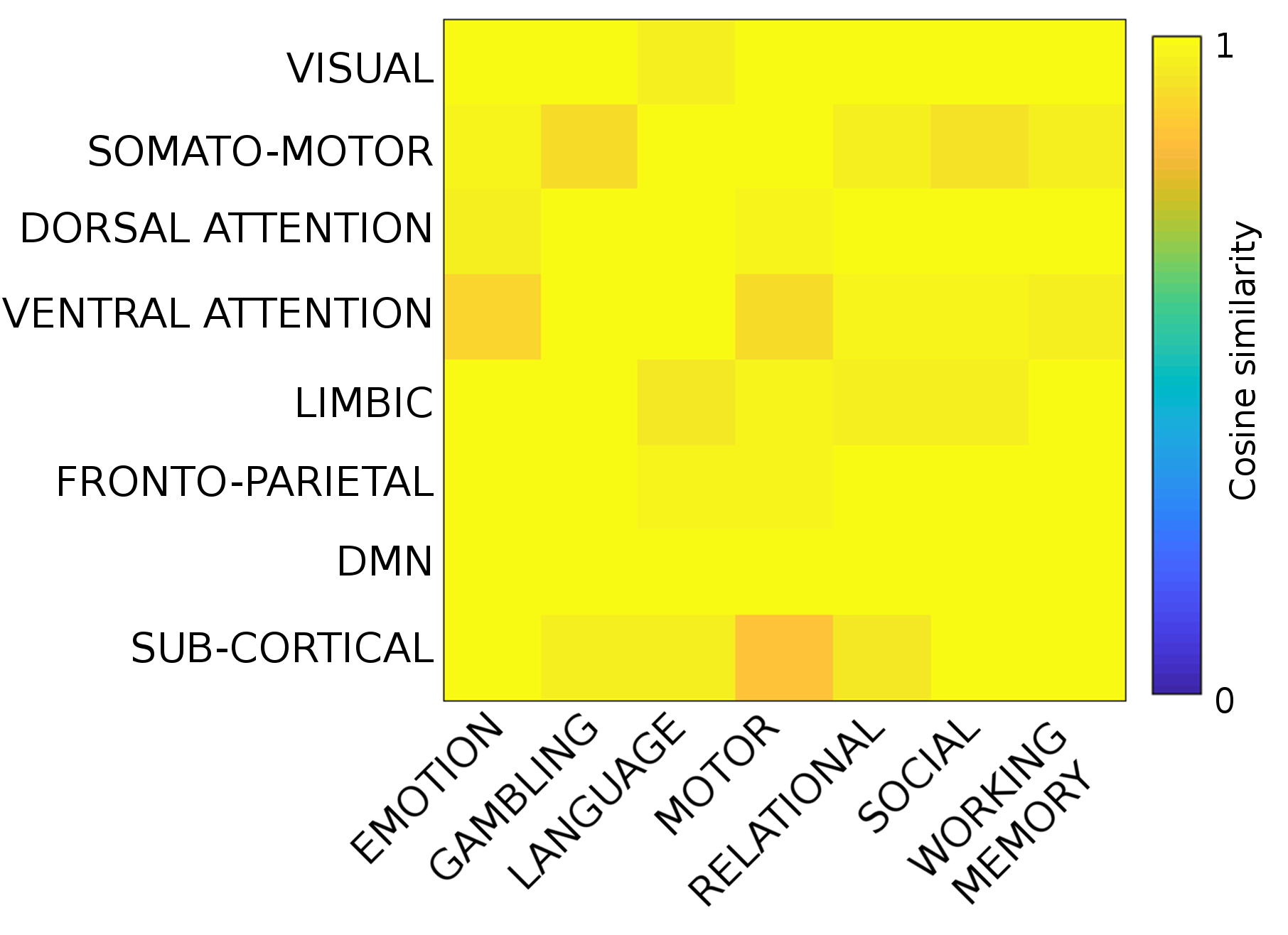}}
\caption{\textbf{Evaluation of the effect of the bandpass filter for REST FCs.}Figure shows the cosine similarity (see \ref{methods}) between pairs of (CP,DP) processing values (see Figure \ref{fig3}) for the two different frequency bands assessed ($[0.001Hz,0.08Hz]$ and $[0.001Hz,0.25Hz]$).
}
\label{figS3}
\end{figure}

\newpage

\begin{figure}[!ht] 
\centerline{\includegraphics[width=1\textwidth]{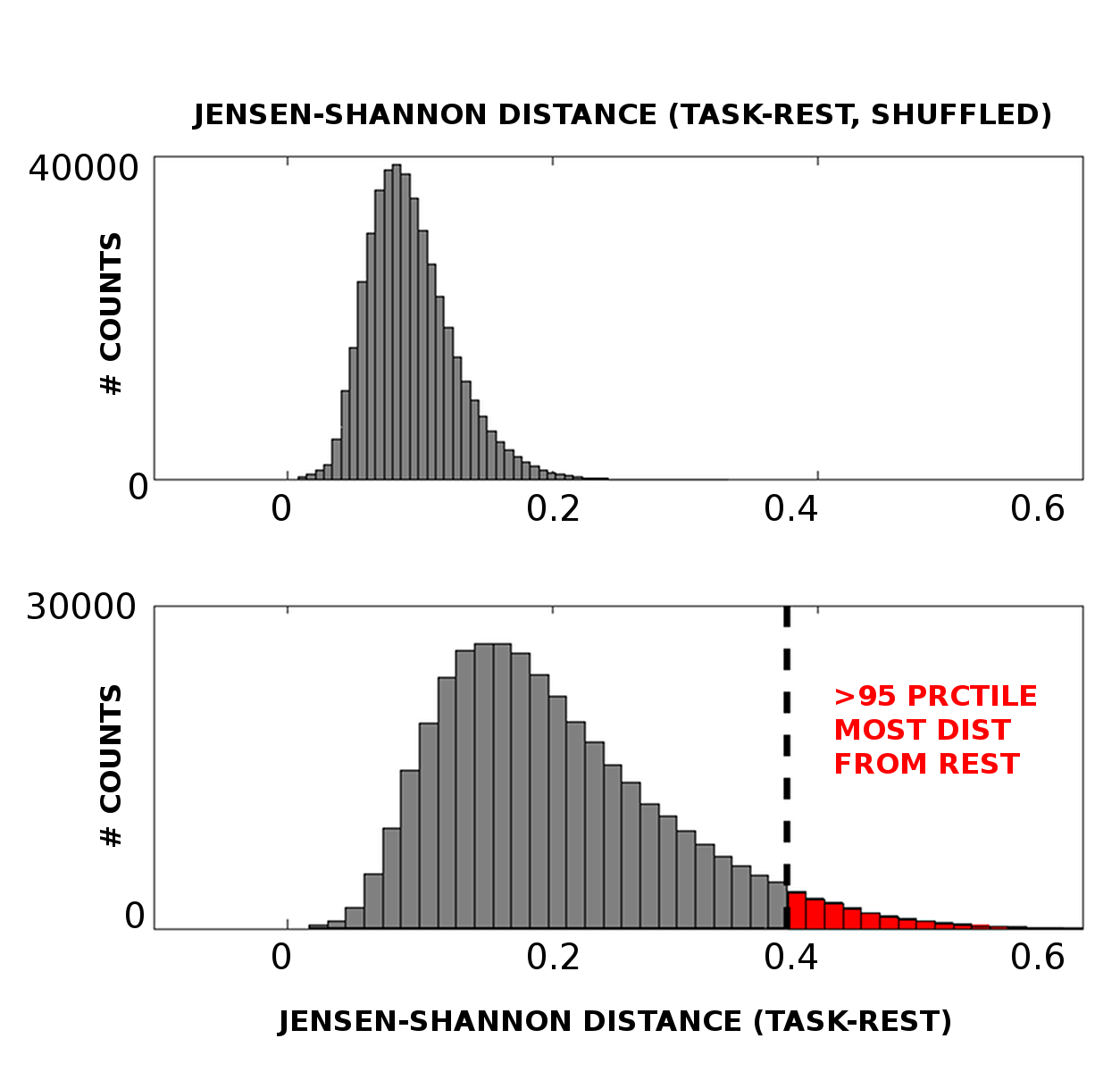}}
\caption{\textbf{Null model for cognitive distance: TASK-REST label permutation test.} Top histogram: distribution of JS distance values when randomly permuting the TASK/REST labels (see \textit{Null model evaluation for connectivity distance analysis} for details). Bottom histogram: actual distribution of Jensen-Shannon distance values across the seven tasks evaluated. The right tail of the histogram is highlighted in red (most distant edges, $> 95th$ percentile), corresponding to the chosen cutoff for centralized and distributed processing evaluation.}
\label{figS4}
\end{figure} 
\newpage


\begin{figure}[!ht] 
\centerline{\includegraphics[width=1\textwidth]{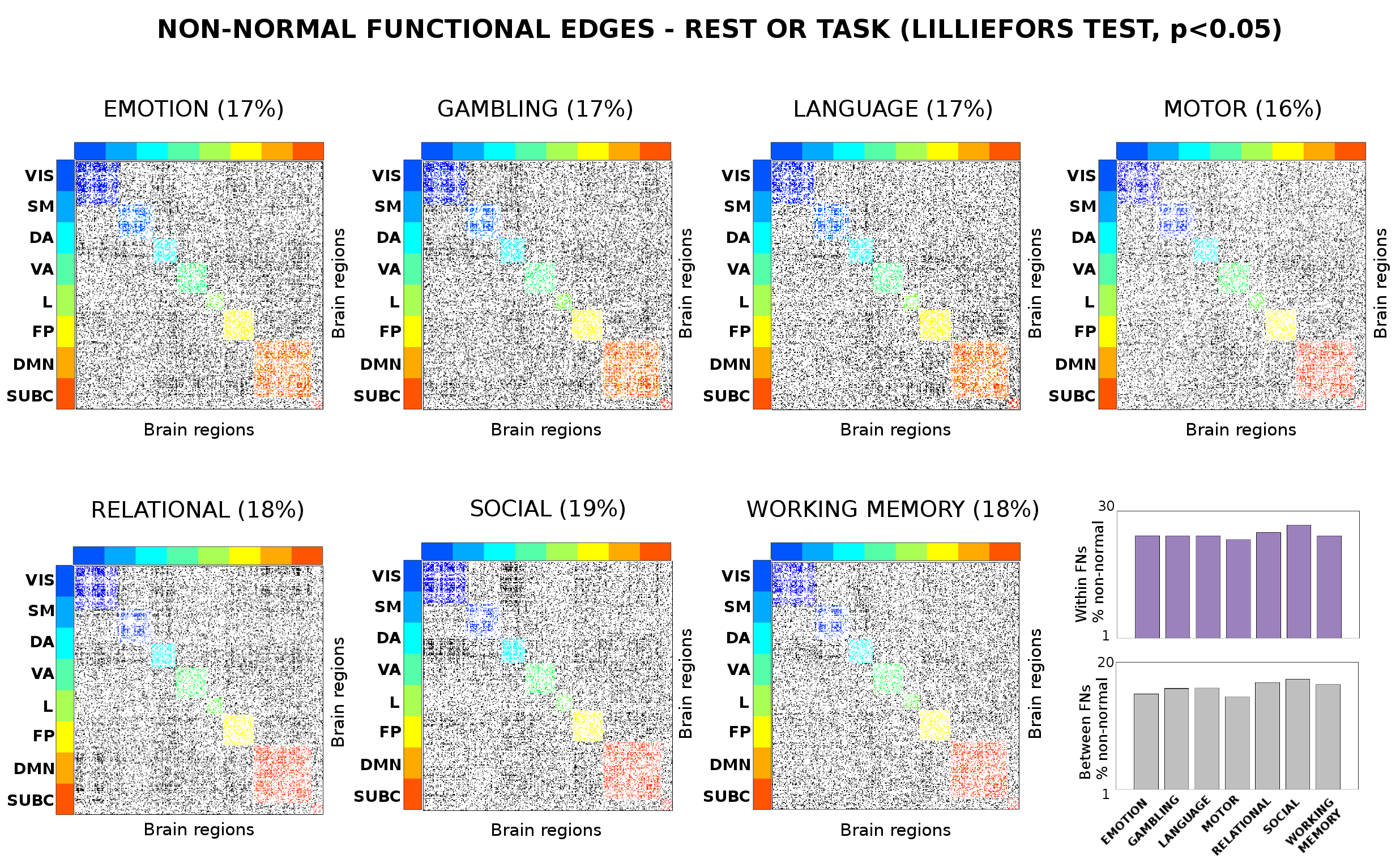}}
\caption{\textbf{Normality test over functional edges.} Results for Lilliefors test, edgewise for each task. Edges that did not pass the normality test (either at rest or on the assessed task) corresponding to within-FN are color-coded accordingly. Edges that did not pass the normality test corresponding to between-FN are depicted in grayscale. The percentage of non-normal edges per task is reported on top of each matrix. The bottom-right bar plots depict the average percentage of within-FNs non-normal edges (violet bars) and the average percentage of between-FNs non-normal edges (grey bars) for each task.
}
\label{figS5}
\end{figure} 
\newpage

\begin{table}[!ht]
\label{tab:table1}
\centering
\begin{tabular}{|p{3cm}||p{3.5cm}|p{3.5cm}|}
\hline
Task name & Most distant edges Surrogate median+99\%  & Most distant edges Original values \\
\hline
Emotion  & 132 $\pm$ 36  & 1655   \\
Gambling  & 18 $\pm$ 14  & 3736   \\
Language  & 8 $\pm$ 9 & 3648   \\
Motor  & 4 $\pm$ 7 & 2625   \\
Relational  & 40 $\pm$ 19 & 4856   \\
Social  & 376 $\pm$ 46 & 3732 \\
Working Memory  & 1 $\pm$ 3 &  4161 \\
\hline
\end{tabular}
\bigskip
\caption{\textbf{Null models for cognitive distance.} Table reports, for each task, the median (and $\pm 99\%$ confidence intervals of the distribution) values for the most distant edges obtained from 100 realization of FC surrogates. The surrogates were built from the original fMRI time series using the Amplitude Adjusted Fourier Transform randomization procedure (\cite{schreiber_surrogate_2000}, see \nameref{surrogates} for details. Note how the original values of most distant edges are always significantly different from the surrogate null distribution (i.e. above the 99 percentile), for all the seven HCP tasks considered in this study.} 

\end{table}

\end{document}